\documentclass[12pt]{article}

\def\be{\begin {equation}}

\def\ee{\end {equation}}

\def\bea{\begin{eqnarray}}

\def\eea{\end{eqnarray}}

\begin{document}

\vspace{5cm}

\thispagestyle{empty}

\begin{center}

{\LARGE \bf Soliton Turbulence as a Thermodynamic Limit of
Stochastic Soliton Lattices}

\vspace{0.5cm}
{\Large Gennady A. El}

Institute of Terrestrial Magnetism, Ionosphere and Radio Wave
Propagation, Russian Academy of Sciences, Troitsk, Moscow Region,
e-mail: el@izmiran.rssi.ru

\vspace{0.5cm}
{\Large Alexander L. Krylov}

O.Yu. Shmidt Institute of Earth Physics, Russian Academy of
Sciences, Moscow

\vspace{0.5cm}
{\Large Stanislav A. Molchanov}

North Carolina University at Charlotte

\vspace{0.5cm}
{\Large Stephanos Venakides}

Duke University, Durham, NC

\end {center}
\vspace{1cm}

{\bf Abstract}
We use recently introduced notion of stochastic soliton lattice
for quantitative description of soliton turbulence.
We consider the stochastic soliton lattice on a special band-gap
scaling of the spectral surface of genus $N$ so that the
integrated density of states remains finite as
$N \to \infty$ (thermodynamic type limit).
We prove existence of the limiting stationary ergodic process and
associate it with the soliton turbulence. The phase space of the
soliton turbulence is a one-dimensional space with the random Poisson
measure. The zero density limit of the soliton turbulence coincides
with the Frish - Lloyd potential of the quantum theory of disordered
systems.

\section{Introduction}

The idea of introducing the statistical description into
the soliton theory has its origin in the work of Zakharov \cite{Z} on
kinetic equation for the rarefied soliton gas. In this work,
a new object, an infinite sequence of KdV solitons on the $x$-axis
was considered. An accurate limit  for the $N$-
soliton solution as $N\to \infty$ is , however, very nontrivial and
has been  investigated so far only for the
case of a very special phase distribution \cite{Gesz}, \cite{Kr} which yields
the reflectionless potential decaying at one of infinities $x\to \pm \infty$.
This kind of infinite - soliton solutions does not provide directly
the spatial uniformity which is necessary for description of the soliton
gas in thermodynamically equilibrium state.

Another view on the problem was proposed by Lax \cite{L1} who noticed
that `the weak limits of  oscillatory sequences of dispersive
compressible flows show a remarkable resemblance to
ensemble averages of classical turbulence theory'. To
constitute such a deterministic analogue of turbulence one has to
construct the correspondent ensemble of flows and provide it with the
physically plausible measure which form a stochastic process.
It is clear from the very
beginning that for description of the `soliton turbulence'
we are interested in the stationary ergodic processes.

The consistent way of application of stochastic description to
the soliton theory was proposed in \cite{EK}, \cite{KE} where
the notion of stochastic soliton lattice
(SSL) was introduced. The basic idea of such a description lies in
the fact that the finite-gap solutions of completely integrable
equations are almost periodic functions and posess, therefore, their
natural stochastic structure \cite{PF} determined by the compact shift
group with the uniform (Haar) measure.

In this work, we investigate $N$-gap stochastic soliton lattices
$q(x)=\nu_N(x)$ of the KdV equation on a special
band-gap
scaling of the spectral Riemann surface of the complex parameter $E$
when the width of gaps is $\sim O(1/N)$ while the
bands are exponentially narrow $\sim \exp(-N)$, $N \gg 1$.
Along with this, the small parameter $1/N$ in our consideration
does not appear explicitely in the Schr\"odinger equation
$(-d^2_{xx}+ q(x))\phi=E\phi$.

The basic results of the present work are:

\begin{itemize}
\item{ the process $\zeta(x)=\lim \limits_{N \to \infty}\nu_N(x)$ exists.
It is an ergodic stationary random process each
realization of which satisfies the KdV equation.
The phase space  of $\zeta(x)$ is a one-dimensional
space with random Poisson measure.}

\item{the considered limit is of a thermodynamic type (the integrated
density of states remains finite as $N \to \infty$ for the chosen
scaling of the band-gap stucture).}

\item{as $ \lim \limits_{N \to \infty} \frac{band}{gap}\to 0$
the thermodynamic limit of stochastic soliton lattices $\zeta(x)$ can be
interpreted as a soliton turbulence.}

\item{ we calculate the rotation number and some ensemble averages for the
soliton turbulence.}

\item{we show that the zero-density limit of the
soliton turbulence yields the Frish -- Lloyd potential of the
quantum theory of disordered systems \cite{FL}, \cite{LGP}.}
\end{itemize}

\section{Basic Preliminaries}

The soliton lattice (SL) ($N$-gap potential
of the Schr\"odinger equation) is given by
the known Its -- Matveev formula (see for instance \cite{DN})

\be \label{sl}
u_N ( x ; {\bf r})= C ({\bf r}) - 2\partial _{xx}^2
\log \Theta [{\bf z (x)}| B({\bf r})] \, ,
\ee

$$
x \in {\bf R}, \qquad {\bf r}= (r_1, \dots , r_{2N+1}), \qquad
r_1<r_2< \dots < r_{2N}< r_{2N+1} \, .
$$
Here
\be \label{theta}
\Theta [{\bf z} | B({\bf r})] = \sum \limits_{{\bf m}} \exp \{
\pi i [2({\bf m},{\bf z}) + ({\bf m},B {\bf m})] \}\, ,
\ee

$$
{\bf m}= (m_1, \dots, m_N ) \in {\bf Z}^N \, , \qquad {\bf z}
\in {\bf C}^N
$$
is the Jacobi theta-function of the hyperelliptic Riemann surface of
genus $N$

\be \label{rs}
\mu ^2 = \prod \limits _{j=1}^{2N+1}(E - r_j) \equiv R_{2N+1}
(E, {\bf r})
\ee
with cuts along the bands.

The Riemann matrix $B({\bf r})$ and the constant $C({\bf r})$ are expressed in
terms of the basis holomorphic differentials $\psi_j$ :

\be \label{B}
B_{ij}= \oint \limits _{\beta_j}\psi_i \, , \ \
C({\bf r})=\sum_{j=1}^{N} r_j
-2\sum_{j=1}^{N}\oint \limits_{\alpha _{j}}E \psi_j \, ,
\ee
Here
\be \label{psi}
\psi_j = \sum \limits _{k=0}^{N-1}a_{jk} \frac{E^{k}}{\sqrt { R _{2N+1}
({\bf r}, E)}}dE \, ,
\ee
and dependence $a_{jk}({\bf r})$ is given by the normalization
\be \label{norm}
\oint \limits_{\alpha _k}\psi _{j} = \delta _{jk}\, ,
\ee
where the $\alpha$-cycles surround the {\it bands} clockwise, and the $\beta$-
cycles are canonically conjugated to $\alpha$'s such that the contour $\beta_j$
starts from the $j$-th cut, then goes to $+\infty$ and returns on the lower cut.

The imaginary phases ${\bf z}(x)$ are given by the formula
\be \label{z}
{\bf z}=-2i {\bf a}_{N-1}x + {\bf d}\, ,
\ee
where {\bf d} is the initial imaginary phase vector.

Using the substitution
\be \label{y}
{\bf z}=\frac{1}{2\pi}B{\bf y}
\ee
we rewrite (~\ref{sl}) in the form with real phases ${\bf y}$ \cite{FFM}
\be \label{uy}
u_N(x|{\bf r})= u_N (y_1(x), \dots, y_N(x)| {\bf r})\, ,
\ee

\be \label{yj}
y_j(x)=k_jx+f_j \, , \qquad f_j (mod 2\pi) \, ,
\ee

\be \label{k}
{\bf k} = -4\pi iB^{-1}{\bf a}_{N-1}\, .
\ee
Here ${\bf k}= {\bf k}({\bf r})= (k_1, \dots, k_N )$
is the wave number vector, and
${\bf f}= (f_1, \dots, f_N) $
is the initial (angle) phase
vector, $-\pi <f_j \leq \pi$.
Note that in (\ref{uy})

\be \label {uq}
u_N(y_1, \dots, y_j+2\pi, \dots, y_N |{\bf r})=
u_N(y_1, \dots, y_j, \dots, y_N |{\bf r})\, ,
\ee
that is $u_N(x|{\bf r})$ is $N$-quasiperiodic in $x$ \cite{N}, \cite{L2}
(hereafter we consider only incommensurate $k_j, j = 1, \dots, N)$.
As for any  almost periodic function, we have for $u_N(x|{\bf r})$ the
Fourier representation

\be \label{F}
u_N(x|{\bf r})= \sum \limits_{j} c_j e^{i(l_jx +h_j)}\, ,
\ee
where $c_j\, , l_j\, , h_j$ are real, $-\pi<h_j\le \pi$.
($l_j \in {\bf M} $, where ${\bf M}$ is a frequency-module \cite{JM}
and $h_j$ are the integer linear  combinations of $f_j (mod 2\pi)$).

The KdV evolution of (\ref{uy}) is isospectral and is described
by the linear motion of the phases on the Jacobian:

\be \label{uyt}
u_N(x,t|{\bf r})=(y_1(x,t), \dots, y_N(x,t)|{\bf r})\, ,
\ee

\be \label{yt}
y_j(x,t)=k_jx+ \omega_j t+f_j \, ,
\ee
where the frequency vector
\be \label{omega}
{\bf \omega}= {\bf \omega}({\bf r}) = (\omega_1, \dots,
\omega_N ) = -8\pi i B^{-1} ({\bf a}_{N-1}\sum \limits_{j=1}^{2N+1}r_j
+ 2{\bf a}_{N-2}) \, .
\ee

\section{Stochastic Soliton Lattices}

It is well known that any almost periodic function generates the
stochastic stationary process (see \cite{JM}, \cite{PF}).

{\bf Definition 1.}\cite{EK}, \cite{KE} The stochastic process
generated by SL $u_N(x|{\bf r})$
we call {\it Stochastic Soliton Lattice} (SSL) and denote as
$\nu _N (x|{\bf r})$.

The general construction of $\nu _N (x|{\bf r})$
admits a very simple and clear description. The realization set
$\Omega$ of it consists of functions (\ref{uy}), (\ref{yj}) where
${\bf f}\in Tor^N$ ; $Tor^N$ is $N$-dimensional torus $(-\pi\, ,\pi]^N$.
The probability measure $d\mu$ is the uniform (Lebesque) measure
on the torus. It corresponds to the description
of $\nu_N (x|{\bf r})$ following from(\ref{uy}),(\ref{yj}):
\be \label{ssl}
\nu_N  (x|{\bf r}) = u_N (\dots k_j x+\phi _j \dots|{\bf r}) \, \
\ee
where $\phi_1, \dots , \phi_N$ are independent random values
uniformly distributed on $(-\pi\, , \pi]$, that is ${\bf \phi}=(\phi_1,
\dots, \phi_N )$ is uniformly distributed on $Tor^N$. As $k_j$ are
incommensurate then $\nu _N (x|{\bf r})$ is an ergodic process \cite{CSF}.

As $\nu _N (x|{\bf r})$ is the stationary process then it has the
Stone - Kolmogorov spectral decomposition \cite{I}; due to the ergodicity
this decomposition has the form (cf. (\ref {F}))

\be \label{SK}
\nu _N (x|{\bf r})= \sum \limits_{j} c_j e^{i(l_jx +\theta_j)}\, ,
\ee
where $\theta _j$ are  uniformly distributed on $(-\pi \, , \pi]$
{\it noncorrelated} random values \cite{PR}.

The well known formula (Bochner - Khintchin) \cite{I} gives us the
covariance function $K(h)$ of the stationary process $\nu _N (x|{\bf r})$:
\be \label{cov}
K(h)\equiv \langle \hat{\nu }_N (x|{\bf r})\cdot
\hat{\nu} _N (x+h|{\bf r})\rangle =
\sum _j |c_j|^2 e^{i\l_j h}\, .
\ee

Here $\hat{\xi}(x) \equiv \xi(x) - \langle \xi \rangle$ is the
centered process.

{\bf Theorem.}  Consider the KdV  equation $u_t-6uu_x+u_{xxx}=0$
as the equation describing the evolution in the phase
space of stationary processes. Let the initial data has the form of the SSL:

\be \label{nu0}
u(x,0)= \nu _N (x|{\bf r}) \, .
\ee
Then the solution of the KdV is
\be \label{nut}
u(x,t)= \nu _N (x|{\bf r})= u(x,0)\, ,
\ee
that is $u(x,0) = \nu_N (x|{\bf r})$ is a stationary point.

{\bf Proof . }The evolution of realizations (\ref{uy}) ,(\ref{yj})
is described by (\ref{uyt}) ,(\ref{yt}). For any moment $t$ one can
introduce the new `initial phase' $f^*_j=\omega _j t +f_j$ which is
also uniformly distributed on $Tor^N$. Therefore, the KdV
evolution changes neither realization set $\Omega$ nor the probability
measure.  Q.E.D.

Now we present some expressions for the ensemble averages
which will be needed in the future.
If $Q(f)$ is an arbitrary smooth function then the mean value $Q(\xi)$
of the stochastic process $\xi$ is

\be \label{aver}
\langle Q(\xi)\rangle = \int \limits_{\Omega} d\mu Q(f) \, , \qquad
f \in \Omega \, .
\ee
(We recall that $\xi(x) \equiv \{\Omega = \{f(x)\}\, , B, \mu \}$,
and $B$ is some $\sigma$-algebra of measurable Borel sets
of $\Omega$).

For $\nu _N (x|{\bf r})$ we have \cite{FFM}
\be \label{ergod}
\langle Q(\nu _N (x|{\bf r})\rangle = \frac{1}{(2\pi)^N}
\int \limits ^{\pi}_{-\pi}\dots \int \limits ^{\pi}_{-\pi}
d\phi_1 \dots d\phi_N Q(\nu _N (x|{\bf r}))=
\lim \limits_{L \to \infty }\frac{1}{L} \int \limits ^{L} _{0}
Q(\nu _N (x|{\bf r})) \, ,
\ee
which constitutes ergodicity of SSL.
 Direct calculation using (\ref{sl}),(\ref{theta}) gives
 surprisingly simple formulas for
 two first moments  of the value $\nu_N (x|{\bf r})$
 Namely,
 \be \label{mom}
 \langle \nu_N (x|{\bf r})\rangle = C({\bf r}) -\frac{1}{\pi i}({\bf k},B{\bf
k})\,
 , \qquad
 \langle \nu_N ^2(x|{\bf r}) \rangle = -
 \frac{1}{3\pi i}({\bf \omega},B{\bf k})\,  ,
 \ee

 and the vectors ${\bf k}({\bf r}), {\bf \omega}({\bf r})$ are given by
 (\ref{k}) , (\ref{omega}).
 It should be noted that expressions (\ref{mom}) are obtained for the
 particular choice of the canonical basis of cycles (see (\ref{norm})).
 However, namely this normalizatiton is preferrable for our consideration.

\section{Rotation Number in Stochastic Soliton Lattices}

Consider the Schr\"odinger equation with an almost periodic
potential $q(x)$
\be \label{schr}
(-\partial ^2 _{xx}+ q(x))\phi=E \phi \, ,
\qquad  x \in {\bf R} \, .
\ee

The potential $q(x)$ has an important characteristics,
{\it the rotation number}, which is defined for real $E$
as [JM]
\be \label{rot}
 \alpha(E)=\lim \limits _{x \to \infty} \frac{1}{x}
 \arg (\phi' (x, E) + i\phi(x,E))\, .
 \ee

We will use also the {\it integrated density of states} ${\cal N}(E)$
which is connected with the rotation number by a 
simple relation (here we improve some inaccuracies in \cite{EK}, \cite{KE}):

\be \label{dens}
{\cal N}(E)= \frac{1}{\pi} \alpha(E) \, .
\ee

If $q(x)$ is N-gap potential , $q(x)=u_N(x|{\bf r})$,
then
\be \label{ap}
\alpha_N(E)=Re \int \limits^{E}_{-\infty} {dp(E')}\, ,
\ee
where $dp(E)$
is the quasimomentum differential \cite{FFM}, \cite{DN}:
\be \label{pnorm} \
dp(E)=\frac{E^N+b_{N-1}E^{N-1}+\dots+b_0}
{\sqrt{R_{2N+1}(E;{\bf r})}}\, ,\qquad
\oint \limits _{\beta_j}dp(E)=0 \, , \qquad j=1,\dots , N \, .
\ee
The integrals of $dp$ over the
$\alpha$ - cycles are known \cite{FFM}, \cite{DN} to give the
components of the wave number vector (\ref{k})

\be \label{kj}
\oint  \limits _{\alpha_j} dp(E)=k_j \, .
\ee

Then,
\bea
&& \frac{1}{2}\sum _{j=1}^{M(E)} k_j \qquad \qquad \qquad
\qquad \hbox{if} \ \ E \in
\hbox{gap}_M \nonumber \\
\alpha_N (E)&=& \label{alpha}  \\
&& \frac{1}{2}\sum _{j=1}^{M(E)} k_j +
\int \limits^{E}_{r_{2M-1}} dp(E') \qquad \hbox{if} \ \ E \in
\hbox{band}_M \, ,\nonumber
\eea
where $M(E)$ is the number of the band nearest to $E$
from the left $M(E) \le N\, , \ M(0)=N\, , \ M(-1)=0$.

We note that the formula (\ref{alpha}) gives an effectivization
for the case of finite-gap potentials of the statement
in \cite{JM} that the values of the function $2\alpha(E)$ if $E \in
\{\hbox{gap}\}$ belong to the frequency module of the almost periodic
potential.

As the SL (\ref{uy}), (\ref{yj}) is the quasiperiodic
function both in $x$ and $t$
one can also formally introduce the {\it  temporal rotation number} 
for the SSL by
\be \label{alphat}
\alpha_N ^t (E)=  Re \int \limits^{E}_{-\infty} {dq(E')}\,
\ee
where $dq(E)$ is the quasienergy differential \cite{FFM},\cite{DN}
($dq(E) =(E^{N+1}+c_N E^{N}+\dots +c_0)dE /\sqrt{R_{2N+1}(E;{\bf r})}$,
$\oint  \limits _{\beta_j}
dq(E)=0\, ,  \ \ \oint  \limits _{\alpha_j}
dq(E)=\omega_j$,
where $\omega _j$ are the frequences (\ref {omega})).

Since the value of $\alpha_N(E)$ is the same for any realization
from the set $\Omega$ from Def.1 then it is the spectral
characteristics of the whole SSL $\nu_N(x|{\bf r})$
(deterministic characteristics of the stochastic process). Due to the
Theorem 1 the rotation number does not change under the KdV evolution.

We introduce also the {\it full integrated density  of states} in the SSL
$\lambda$ which is a number:
\be \label{lambda}
\lambda \equiv  {\cal N}_N(\infty)=
\frac{1}{2\pi}\sum _{j=1}^N k_j \, .
\ee
One can see that the full integrated density of states has in the SSL
the natural meaning of the mean number of waves per unit length.

\section{Thermodynamic Limit of Stochastic Soliton
Lattices}
Let the nontrivial (band-gap) part of the spectrum lies in the interval
$(-1, 0)$ of the real $E$-axis: $\alpha_N(-1)=0\, ,\ \alpha_N(0)=\pi \lambda$.
For $N\gg1$ we consider the following scaling of the band-gap structure
\be
\label{scal} \hbox{gaps}(E) \sim \frac{1}{N} \, \qquad \hbox{bands}(E) \sim
\exp{(-N)}\, \qquad N\gg1 \, , \ \  E \in (-1,0)\, .
\ee
Using (\ref{k}) one can show that the scaling (\ref{scal}) implies
the following behavior for the wave numbers: $k_j \sim 1/N$, which
provides  finiteness of the rotation number $\alpha_N(E)$
(and of the integrated density of states ${\cal N}_N(E)$) as
$N \to \infty$:
\be \label{aturb}
\alpha(E) \equiv
\lim \limits_{N \to \infty} \alpha_N(E)=
\lim \limits_{N \to \infty}\frac{1}{2}\sum _{j=1}^{M(E)} k_j < \infty
\, ,\qquad  M \le N \,.
\ee
Due to this property we call the scaling (\ref{scal}) and the
correspondent limit as $N \to \infty$ the {\it thermodynamic} ones.

It should be noted that the thermodynamic scaling (\ref{scal})
appears when one considers the quasiclassical asymptotics of the
spectrum for periodic potentials \cite{WK}, \cite{V1},
\cite{V2}. In contrast to
indicated works, however,   the small parameter $1/N$ in
our consideration is contained only in the band-gap
structure of the spectrum and does not appear explicitely in the
Schr\"odinger equation (\ref{schr}).  As a result, we will show that
the limit for the finite-gap potentials on the thermodynamic
scaling exists as $N \to \infty$ {\it in a strong sense}.
We emphasize the difference of this point from the known
Lax -- Levermore -- Venakides (LLV) approach (see \cite{LL},
\cite{V1}, \cite{V2}) to the limiting
passage $\lim_{N \to \infty} u_N$ where due to rescaling $x \to Nx$ the above
limit exists only in a weak sense.  The LLV consideration provides information
about the slow modulations ({\it macrostructure}) appearing in the $N$-soliton
(or $N$-gap) KdV solution under the time evolution.  We, on the contrary, are
interested in the limiting {\it microstructure} of nonmodulated
$N$-phase wave as $N \to \infty$.  Futhermore, we are going
to show that the thermodynamic limit exists
for the whole stochastic process (SSL) which implies not only
compactness of
the realization set but also convergence of the probability measure.

Now we describe  the thermodynamic scaling more in detail.
Following \cite{V2} we introduce the lattice of points
\be \label{lat}
1\approx \eta _1>\eta _2>\ldots >\eta _N\approx 0\, ,
\ee
where
$$
-\eta _j^2=\frac 12\left( r_{2j-1}+r_{2j}\right)\,
$$
are centers of bands.

We define two continuous functions on the lattice:

1. The normalized density of bands: $\varphi(\eta)d\eta \approx
\frac{\hbox{number of bands in} \  (E,E+dE)}{N}$
\be \label{phi}
\varphi(\eta_j)=
\frac{1}{N(\eta_j - \eta_{j+1})} +O(\frac{1}{N})\, ,
\qquad \int \limits_0^1
\varphi(\eta)d\eta =1\, , \ \ \eta^2 = -E \in (0, 1) \, .
\ee

2. The normalized logarithmic band width
\be\label{gam}
\gamma(\eta_j)= -\frac{1}{N} \log(r_{2j}-r_{2j-1})+ O(\frac{1}{N})\,.
\ee

Now one can easily  represent the relationship (\ref{aturb}) for the
thermodynamic limit of the rotation number in a continuum
form:
\be \label{conta}
2d\alpha(-\eta^2) = \varphi(\eta)k(\eta) d\eta \, .
\ee
where $k(\eta)$ is the continuãü äøüøå of the
wave number vector (\ref{k}).

For the temporal rotation number $\alpha_N^t (E)$
(\ref{alphat}) in the thermodynamic limit we have the equation
which is analogous to (\ref{conta}):

\be \label{contat}
2 d\alpha^t(-\eta^2) = \varphi(\eta)\omega(\eta) d\eta \, .
\ee
where $\omega(\eta)$ is the continuum limit for the frequency vector
(\ref{omega}).

To find a continuum limit for the wave number and the frequency
vectors given by (\ref{k}) and (\ref{omega}) one
has to know the continuum limits of the period matrix $B$ (\ref{B})
and of the
vectors  ${\bf a}_{N-1}$ and ${\bf a}_{N-2}$ from (\ref{psi}).
It is clear that all quantities given on the Riemann surface
can be computed now in
terms of two functions (densities) $\varphi(\eta)$ and $\gamma(\eta)$
completely describing the band - gap structure.

In particular, the period matrix
(\ref{B}) and the vectors ${\bf a}_{N-1}$ and
$({\bf a}_{N-1}\sum \limits_{j=1}^{2N+1}r_j + 2{\bf a}_{N-2})$
take the form \cite{V2}:

\be \label{cB}
B \to B(\eta, \mu)= -\frac{i}{\pi}\left(\log \left|\frac{\eta - \mu}
{\eta+\mu}\right| \varphi(\mu)+ \gamma(\mu) \delta(\eta-\mu) \right)\, ,
\ee
where $\delta (x)$ is the Dirac delta-function;
\be \label{ca}
{\bf a}_{N-1} \to -\frac{\eta}{2\pi } \, , \ \
{\bf a}_{N-1}\sum \limits_{j=1}^{2N+1}r_j + 2{\bf a}_{N-2} \to
\frac{2\eta^3}{\pi }\, .
\ee

Now we represent the general relationship (\ref{k}) between the wave number
vector ${\bf k}$ and the vector  ${\bf a}_{N-1}$ in the form
suitable for the limiting transition
\be \label{kB}
{\bf k}B= -4\pi i {\bf a}_{N-1}
\ee
Then applying (\ref{conta}), (\ref{cB}), (\ref{ca}) to (\ref{kB})
we arrive at the integral equation for the rotation number in the
thermodynamic limit :

\be \label{inta}
\frac{1}{\pi}\varphi(\eta)d\eta \int\limits_0^1 \log \left|\frac{\eta - \mu}
{\eta+\mu}\right| d\alpha(-\mu^2)  + \gamma(\eta) d\alpha(-\eta^2)
=-\eta \varphi(\eta)d\eta
\ee

Similarly, for the thermodynamic limit of the temporal
rotation number one gets with the aid of  (\ref{contat}), (\ref{cB}),
(\ref{ca}) and (\ref{omega}):
\be \label{intat}
\frac{1}{\pi}\varphi(\eta)d\eta
\int\limits_0^1 \log \left|\frac{\eta - \mu} {\eta+\mu}\right|
d\alpha^t(-\mu^2) + \gamma(\eta) d\alpha^t(-\eta^2) =
4\eta^3 \varphi(\eta)d\eta \, .
\ee

Thus, we have established the existence of the thermodynamic limit
for the rotation number in SSL (as a matter of fact, the rotation number,
as well as in the finite-gap case, is a deterministic function).

The rotation number is known to be an imaginary part of the
Floquet exponent $w(E)$ (for the finite-gap potentials
$w(E)=i\int dp(E)$) \cite{JM}. Then we have with the aid of
Herglotz formula:
\be \label{herg}
w(E)=\int \limits^{\infty}_{-\infty}\frac{\alpha(z)dz}{E-z}\, ,
\qquad Im E>0 \, .
\ee
But following Kotani \cite{Kot}, if the Floquet exponent satisfies the
conditions
\bea
&& Re w(E+i0)=0 \ \ \hbox{a.e.} \ \ \hbox{on} \
[0, +\infty) \label{prop} \\
&& \int \limits^{0}_{-\infty}-Re w(E+i0)d\alpha(E)=0 \nonumber\, ,
\eea
then there exists an ergodic stationary random process whose
Floquet exponent is $w(E)$.
One can see that for SSL conditions (\ref{prop})
are satisfied due to the properties of the quasimomentum
(\ref{pnorm}). Therefore   the thermodynamic limit of SSL exists.

{\bf Remark.} It can be shown that the existence of the limit for
the rotation number guarantees the existence of the limit for the
Weyl function. Then the compactness of the realization set can
be deduced from the results of Marchenko \cite{M}.

As $band/gap \to 0$ in the thermodynamic limit (see (\ref{scal})),
then it is natural to call the thermodynamic limit $\zeta(x)$
of the SSL $\nu_N(x)$ a {\it soliton turbulence}
\be \label{turb}
\zeta(x) = \lim \limits_{N \to \infty} \nu_N(x) \, .
\ee
We believe that the spectrum of a typical realization of the soliton
turbulence is pure point on $(-1, 0)$ but exact proof of this conjecture
is missed yet.

The ensemble averages  (moments) (\ref{aver}) are the important
characteristics of the soliton turbulence. To compute three first moments
we make use of the expressions (\ref{mom}) obtained for the finite-gap SSL.
These expressions admit direct limiting passage as $N \to \infty$ in the
thermodynamic scaling (\ref{scal}).

Using results of \cite{V2} it is not hard to show that the constant
$C({\bf r})$ (\ref{B}) in the Its-Matveev formula (\ref{sl}) vanishes
in the thermodynamic limit. Then, substituting (\ref{conta}),
(\ref{contat}) into (\ref{mom}) we have for the two first moments
in soliton turbulence:
\be \label{mom1}
\langle \zeta \rangle =-2\lambda \int \limits^0_{-1}d\tilde \alpha(E)=
-2\lambda\, ,
\ee

\be \label{mom2}
\langle \zeta^2 \rangle =\frac{4\lambda}{3}\int \limits^0_{-1}Ed\tilde
\alpha(E)= -\frac{4\lambda}{3}\overline{E}\, .
\ee
Here
$$
d\tilde
\alpha(E) = \frac{d\alpha}{\pi \lambda}\, , \qquad
\int\limits_{-1}^0 d\tilde\alpha = 1\, , \qquad \overline{f(E)}
= \int\limits_{-1}^0 f(E)d\tilde\alpha(E)\, .
$$
Thus, we have expressed the ensemble averages in soliton turbulence
through the averages over the spectrum with the measure
$d\tilde \alpha(E)$. One should also note that the Floquet exponent
$w(E)$ (\ref{herg}) provides the averaged Kruskal integrals as the
coefficients in the decomposition as $E \to \infty$.

\section{Poissonic Properties of Soliton Turbulence}

We study the phase space of the soliton turbulence. We recall
that in the finite-gap SSL the phase space is the $N$ - dimensional
torus with the uniform (Lebesque) measure. For  performing the
thermodynamic limit it is convenient to introduce the linear
phases instead of the angle phases $\phi_j$ in the SSL (\ref{ssl})

\be \label {l}
l_j \equiv \frac {\phi_j}{k_j}\, ,
\ee
where $l_j \, \ (j=1, \dots, N)$ are independent random values uniformly
distributed on $(-\frac{\pi}{k_j}, \frac{\pi}{k_j}]$ respectively.

The further consideration is close to standard procedure of the
thermodynamic limiting passage in the ergodic theory of
ideal gas (see for instance \cite{S}, \cite{CSF}). We now turn to
a new space with the aid of factorization of the torus by
interchange group $S_N$.
The sets of $N$ points $l_j$ serve now as the points of the obtained space
$Q_N=Tor^N/S_N$. The measurable sets in the factorized space
consist of realizations having $0, 1, \dots, N$ linear phases
in the interval $\Delta$  of the $x$ axis.

We introduce the random value $\xi_j = \chi _{(0,1)}(l_j)$
which is the number of hitting
of the particular phase $l_j$ into the fixed interval
$(0,1) \subset {\bf R}$
(we suppose that $\pi /{k_j}\geq 1$).
The variable $\xi _j$ takes two values:
$1$ and $0$ with the probabilities $p_j(1)= k_j/{2\pi}$ and
$p_j(0)=q_j=1-p_j=1-{k_j}/{2\pi}$ .
The generating function $\varphi _j(z)$ for $\xi_j$ is

\be \label{phij}
\varphi _j(z)=(1-p_j) + zp_j
\ee

The sum $\xi^{(N)}\equiv \sum \limits _{j=1}^N \xi_j$  is the number of
hitting of all linear phases into $(0,1)$. As $\xi_j$ are independent
random values,
then the generating function for $\xi^{(N)}$ has the form

\be \label{phig}
\varphi^{(N)}(z)= \prod _{j=1}^{N}\varphi _j(z)= \prod _{j=1}^{N}
(1+(z-1)p_j)=\prod _{j=1}^{N}(1+\frac{(z-1)k_j}{2\pi}) \, .
\ee

On the thermodynamic scaling (\ref{scal}) $k_j=O(N^{-1})$.
Then taking the logarithm of (\ref{phig}) we obtain

$$
\ln \varphi^{(N)}(z)= (z-1)\frac{1}{2\pi}\sum _{j=1}^{N}k_j + O(N^{-1}) \, .
$$

Therefore taking into account (\ref{lambda}) one has
\be \label{phii}
\varphi^{(\infty)}(z)=\exp\{(z-1)\lambda \} = \sum \limits^{\infty}_{n=0}
z^n \left( \frac{e^{-\lambda}\lambda^n}{n!} \right) \, .
\ee

Thus, the right-hand
part of (\ref{phii}) is the generating function for
the {\it Poisson distribution} with the parameter $\lambda$
( which is the full integrated density of states in our case).
Therefore, the limiting measure in the configurational space
is the random Poisson measure (Poissonic white noise).

We note that the Poissonic white noise can be described as the
random collection of the points $\{l_j\}$ (the linear phases
in our case) on the $x$ - axis such that the distances $s_k$
between them are independent random values distributed exponentially
with the density
\be \label{f}
f(s)=\lambda \exp(-\lambda s)\, .
\ee
We also note that the Poisson parameter $\lambda$  according to
(\ref{mom1}) can be expressed through the first moment in
the soliton turbulence:
$$
\lambda = -\frac{\langle \zeta \rangle}{2} \, .
$$
The value $\lambda$ can be then interpreted as a density of
the soliton turbulence.

\section{The Frish -- Lloyd Potential as a Zero-Density
Limit of the Soliton Turbulence}

We consider the soliton turbulence with the small integrated density
of states. We introduce a small parameter $\varepsilon \ll 1$ and
make a renormalization:
\be \label{renorm}
\alpha(E)=\varepsilon \tilde\alpha(E) \, , \qquad
\lambda=\varepsilon \tilde\lambda \, .
\ee

This implies the following transformations for the functions
$\varphi(\eta)$ and $\gamma(\eta)$ (\ref{phi}), (\ref{gam})
characterizing the Riemann surface in the thermodynamic scaling
\be \label{ren}
\varphi(\eta)=\varphi(\eta)\,, \qquad \gamma(\eta) =
\frac{\tilde\gamma(\eta)}{\varepsilon} \, , \qquad \tilde\gamma(\eta) =
O(1)\, .
\ee

We choose $\tilde\gamma(\eta)= c\eta$ which corresponds to imposing
the periodicity condition :
$\forall k_j=2\pi c ^{-1}(\varepsilon N^{-1})$.
Then, it follows from (\ref{lambda}), (\ref{renorm}) that
$\tilde \lambda = 1/c$.

We also make rescaling of the spatial variables
\be \label{spat}
x=\frac{\tilde x}{\varepsilon}\, , \qquad l_j=-\frac{\tilde l_j}
{\varepsilon}\, .
\ee

Consider now the SSL with a small integrated density of states
on the thermodynamic scaling (\ref{scal}). We make use of the
Its -- Matveev formula (\ref{sl}):
\be  \label{sl1}
\nu_N(\tilde x)\approx -2\varepsilon^2\partial _{\tilde x \tilde x}^2
\log \Theta_N (\dots, \frac{-2ia_{N-1, j}(\tilde x- \tilde l_j)}
{\varepsilon}, \dots| B) \, ,
\ee
where according to (\ref{cB}), (\ref{ca})
\be \label{cB1}
B_{jj}\approx \frac{\tilde\gamma(\eta_j)}{\varepsilon}\, ,
\qquad B_{ij} \approx \log\left|\frac{\eta_i-\eta_j}{\eta_i+\eta_j}
\right| \, ,
\ee
 \be \label{ca1}
a_{N-1, j} \approx -\frac{\eta_j}{2\pi} \, .
\ee
An analysis of the explicit expression for the theta-function
(\ref{theta}) provided (\ref{cB1}), (\ref{ca1}) shows that on the
set of realizations excluding the set of a small measure
one can
neglect the contribution of the off-diagonal part of the period
matrix $B$ into
the solution (\ref{sl1}) (one suppose the inequality \be \label{ineq} \sum
\limits^N_{j=1} \frac{\tilde\gamma(\eta_j)}{\varepsilon} n_j^2 \gg  \sum
\limits^N_{i\ne j=1}  \log\left|\frac{\eta_i-\eta_j} {\eta_i+\eta_j}
\right| n_i
n_j   \, , \qquad N\gg 1 \, .  \ee to be satisfied outside of the
indicated set of a small measure).

Then asymptotically in $N$, $\varepsilon$,
($N\gg1\, \ \varepsilon \ll 1\, ,\ N\varepsilon \gg 1$ )
the following factorization is valid :
\be \label{fact}
\Theta_N \approx \prod \limits^N_{j=1} \Theta(\eta_j
\frac{(\tilde x- \tilde l_j)}{\pi\varepsilon}|\tau_j)\, ,
\ee
where $\Theta(y)$ is the one-dimensional theta-function
($\Theta_3$ in standard literature (see for ex. \cite{AS})),
\be \label{tau}
\tau_j=\frac{K(1-m_j)}{K(m_j)} \approx 0 \, , \qquad
m_j \approx 1-\frac{1}{\eta_j}\exp(-\frac{c}{\varepsilon}\eta_j)
\approx 1 \, .
\ee
Here $K(m)$ is the complete elliptic integral of the first kind,
$m_j$ are the ellipticity parameters (do not confuse with the
multiintegers in (\ref{theta}), (\ref{ineq}) ); $m_j=1$ corresponds
to the soliton limit.

Consider the rescaled SSL of a small density
$\nu_N(\tilde x)/\varepsilon^2$.
Then it follows from (\ref{sl1}), (\ref{fact}), (\ref{tau}) that
\be \label{}
\frac{\nu_N(\tilde x)}{\varepsilon^2}
\approx -2 \sum \limits^N_{j=1}\eta_j^2 \frac{sech ^2\left(
\eta_j\frac{\tilde x - \tilde l_j}
{\pi\varepsilon}\right)}{\varepsilon^2}\, .
\ee

Then passing to a limit
\be \label{lim}
\varepsilon \to 0 \,, \qquad N\to \infty \, , \qquad
\varepsilon N \to \infty \,
\ee
one obtains
\be \label{FL}
\tilde\zeta(\tilde x) = \lim \frac{\nu_N(\tilde x)}{\varepsilon^2}=
-4\sum \limits_j a_j^2 \delta(\tilde x - \tilde l_j) \, ,
\ee
where, according to the results of Sec. 5, $\tilde l_j$ are
random points on the $\tilde x$ - axis. The distances $\tilde s_k$
between these points are independent random values distributed with the
density
\be \label{f1}
f(\tilde s)=\frac{1}{c} \exp(- \frac{\tilde s}{c})\, .
\ee
The amplitudes $a_j$ are independent among themselves and independent
of $\tilde l_j$ random values  distributed with the density
$\varphi(a)$.
The distribution (\ref{FL}) is so-called complex Poisson white noise;
it is called the Frish -- Lloyd potential in the quantum theory
of disordered systems \cite{FL}, \cite{LGP}.

\vspace{0.5cm}
{\bf Aknowledgements}

We are grateful to M. Freidlin, P.Miller and S.Novikov for useful discussions.

\end{document}